\documentclass[twocolumn,prb,showpacs]{revtex4}
\usepackage{amsmath,amsthm}
\usepackage{epsfig}
\usepackage{graphics}
\usepackage{graphicx}
\usepackage{dcolumn}
\usepackage{bm}

\begin{document}

\title{Electrodynamics of the vanadium oxides VO$_2$ and V$_2$O$_3$}
\author{M. M. Qazilbash,$^{1,\ast}$ A. A. Schafgans,$^1$ K. S.
Burch,$^{1,\dag}$ \\S. J. Yun,$^2$ B. G. Chae,$^2$ B. J. Kim,$^2$
H. T. Kim,$^2$ and D. N. Basov$^1$}
\affiliation{$^{1}$Physics Department, University of California-San Diego, La Jolla, California 92093, USA \\
$^{2}$IT Convergence and Components Lab, ETRI, Daejeon 305-350,
Korea}

\date{\today}

\begin{abstract}
The optical/infrared properties of films of vanadium dioxide
(VO$_2$) and vanadium sesquioxide (V$_2$O$_3$) have been
investigated via ellipsometry and near-normal incidence
reflectance measurements from far infrared to ultraviolet
frequencies. Significant changes occur in the optical conductivity
of both VO$_2$ and V$_2$O$_3$ across the metal-insulator
transitions at least up to (and possibly beyond) 6 eV. We argue
that such changes in optical conductivity and electronic spectral
weight over a broad frequency range is evidence of the important
role of electronic correlations to the metal-insulator transitions
in both of these vanadium oxides. We observe a sharp optical
transition with possible final state (exciton) effects in the
insulating phase of VO$_2$. This sharp optical transition occurs
between narrow $a_{1g}$ bands that arise from the
quasi-one-dimensional chains of vanadium dimers. Electronic
correlations in the metallic phases of both VO$_2$ and V$_2$O$_3$
lead to reduction of the kinetic energy of the charge carriers
compared to band theory values, with paramagnetic metallic
V$_2$O$_3$ showing evidence of stronger correlations compared to
rutile metallic VO$_2$.
\end{abstract}


\maketitle

\section{Introduction}

Vanadium dioxide (VO$_2$) and vanadium sesquioxide (V$_2$O$_3$)
have been subjects of intensive experimental and theoretical
studies for the past several decades. These systems are considered
classical prototypes of materials exhibiting metal-insulator
transitions.\cite{mottbook,imada-rmp} An interesting and
intriguing aspect is that the electronic metal-insulator
transitions in VO$_2$ and V$_2$O$_3$ are accompanied by changes in
the lattice structure. While there has been progress towards
understanding the roles played by electron-electron interactions
and the lattice distortion in the metal-insulator transitions,
there is no consensus on the mechanism of the metal-insulator
transitions, at least in the case of VO$_2$. An important
difference between the two oxides is that V$_2$O$_3$ is
anti-ferromagnetically ordered in the insulating phase whereas the
most stable insulating phase ($M_1$) of VO$_2$ does not exhibit
magnetic ordering. In $M_1$ VO$_2$, the Peierls distortion leads
to dimerization (pairing/charge-ordering) of vanadium
ions~\cite{goodenough} thereby preventing magnetic ordering in an
otherwise correlated insulator.\cite{mottbook,mottprb} Another
difference between these two oxides is that the nominal valence of
vanadium ions in VO$_2$ is +4 whereas in V$_2$O$_3$ it is +3.

VO$_2$ undergoes a metal-insulator transition (MIT) at $T \approx$
340 K from a low temperature insulating phase to a high
temperature rutile metallic phase. The resistivity decreases by
four orders of magnitude across the MIT accompanied by a
structural transition from the monoclinic unit cell in the
insulator ($M_1$) to a tetragonal unit cell in the rutile metal
($R$). In VO$_2$, the vanadium ions are in the +4 valence state,
and therefore a single electron resides in the $d$-orbitals, a
scenario supported by theoretical calculations and by recent
experiments.\cite{allenprl,eyert,georges,softxray,xray-photo}
There has been much controversy on the driving mechanism of the
MIT in VO$_2$, and over the relative importance of the Peierls
scenario within the single particle picture and electronic
correlations representing Mott
physics.\cite{allenprl,eyert,georges,softxray,xray-photo,pougetnmr,
pougetprl,gupta,paquet,ricecomment,GW,kimnewjphys,laad,kimprl,lupi,averitt}
We have recently demonstrated divergent effective quasiparticle
mass in VO$_2$ as evidence that the MIT is a Mott
transition.\cite{massdivergence,brinkman,kim} There is, however, a
structural component to the phase transition in the form of a
Peierls instability (charge density wave) in the monoclinic
insulating ($M_1$) phase of VO$_2$ which leads to unit cell
doubling and the formation of vanadium dimers (pairs) along the
$c$-axis (in the rutile basis). The presence of such vanadium
chains imparts a quasi-one-dimensional character to what is
essentially a three dimensional system. The competing effect of
the Peierls instability prevents long-range magnetic ordering in
the correlated $M_1$ phase of VO$_2$. Therefore, the $M_1$ phase
of VO$_2$ should be classified as a Mott insulator that is
charge-ordered and not magnetically-ordered.\cite{massdivergence}

V$_2$O$_3$ undergoes a first-order metal-insulator transition at
$T$ $\approx$ 150 K from a low temperature antiferromagnetic
insulating (AFI) phase to a high temperature paramagnetic metallic
(PMM) phase. The crystal structure also deforms from monoclinic in
the insulating phase to rhombohedral symmetry in the metallic
phase. The V-V distance along the c-axis (in hexagonal basis) is
larger in the monoclinic insulating phase compared to the
rhombohedral metal.\cite{dernier} While there is debate on the
significance of cation-cation covalent bonding along the $c$-axis
to the physics of
V$_2$O$_3$,\cite{castellani,ezhov,mila,joshi,matteo,tanaka} the
change in crystal structure of V$_2$O$_3$ is not associated with
Peierls instability and unit-cell doubling, and magnetic ordering
is preserved in the insulating phase. From charge balancing in the
chemical formula of V$_2$O$_3$, the vanadium ions are in the +3
valence state and two electrons are present in the $d$-orbitals of
vanadium. Despite extensive experimental and theoretical effort,
there is no consensus on precisely how the two electrons are
distributed within the $t_{2g}$ manifold of the
$d$-levels.\cite{castellani,ezhov,mila,joshi,matteo,tanaka,bao,paolasini,
park,mo} There is, however, an agreement that the single-band
Hubbard model is inadequate to describe the MIT in V$_2$O$_3$
given the complex multi-orbital character of this material. The
experimental results reported here provide further constraints on
the intrinsic electronic structure of the PMM and AFI phases of
V$_2$O$_3$.

Given that both of these oxides of vanadium exhibit
metal-insulator transitions that are accompanied by changes in
lattice structure, it is important to understand, on the same
footing, their optical properties across the phase transition. A
comprehensive study of the optical properties of VO$_2$ and
V$_2$O$_3$ will help us gain insight into how charge-ordering in
insulating VO$_2$ and magnetic ordering in insulating V$_2$O$_3$
influence the electrodynamics of their respective insulating
phases and the neighboring metallic phases. There have been
several previous studies that have addressed the optical
properties of VO$_2$ and/or
V$_2$O$_3$.\cite{barker,verleur,shin,thomasjltp,thomasprl,rozenbergprl,rozenbergprb,nohprb,okazaki}
However, none of these studies have reported optical constants of
both VO$_2$ and V$_2$O$_3$ over the wide frequency (far infrared
to ultraviolet (6 eV)) and temperature ranges that we have been
able to investigate in this work.

We find that VO$_2$ and V$_2$O$_3$ are correlated materials in
which the important energy scale is the Hubbard intra-atomic
Coulomb repulsion energy $U$ leading to changes in optical
conductivity and spectral weight at least up to 6 eV. We find that
$M_1$ VO$_2$, a Mott insulator that is charge-ordered, exhibits a
sharp, distinct feature in the optical conductivity that likely
results from the quasi-one-dimensional character of the density of
states related to the one-dimensional chains formed by vanadium
pairs. We discuss the possibility that this feature in the optical
conductivity of VO$_2$ also has an excitonic origin due to final
state effects. Such a distinct feature is absent in the optical
conductivity of the Mott insulator V$_2$O$_3$ which is
magnetically ordered and lacks the quasi-one-dimensional character
associated with charge-ordering. Charge-ordering in insulating
VO$_2$ may partially circumvent the effects of the intra-atomic
Coulomb repulsion or Hubbard $U$.~\cite{jan} Nevertheless, the
intra-atomic Coulomb repulsion is necessary for opening the energy
gap in $M_1$ VO$_2$ as well as in AFI V$_2$O$_3$. Moreover,
electronic correlations dominate charge dynamics in the metallic
phases of both VO$_2$ and V$_2$O$_3$, with PMM V$_2$O$_3$ more
strongly correlated compared to rutile VO$_2$.

\section{Experimental Methods}

Previous optical and infrared studies on V$_2$O$_3$ crystals
addressed the issue of changes in spectral weight across the MIT
in the energy range up to $\approx$ 1.5
eV.\cite{thomasjltp,thomasprl,rozenbergprl,rozenbergprb} The
availability of high quality V$_2$O$_3$ and VO$_2$ films, and
recent advances in broad band ellipsometry at cryogenic and
elevated temperatures enable accurate determination of optical
constants and changes in spectral weight with temperature over a
wide energy (frequency) range.\cite{vdmarel} This is because
ellipsometry is a self referencing measurement and the complex
ellipsometric constants (real part $\Psi$($\omega$) and imaginary
part $\Delta$($\omega$)) can be obtained at each incident
frequency thereby precluding the need for Kramers-Kronig
transformations.\cite{tompkins} The MIT is literally destructive
for pure VO$_2$ and V$_2$O$_3$ crystals which could make accurate
measurements in both metallic and insulating states on the same
sample very challenging. The VO$_2$ and V$_2$O$_3$ films, on the
other hand, do not show signs of deterioration even after going
through several cycles across the MIT.\cite{footnote1}

VO$_2$ films about 100 nm thick were grown on ($\bar{1}$012)
oriented sapphire (Al$_2$O$_3$) substrates by the sol gel method
and the details of growth and characterization are given
elsewhere.\cite{chae} Films of V$_2$O$_3$ are 75 nm thick and are
grown on (10$\bar{1}$0) oriented sapphire substrates. The
V$_2$O$_3$ films are obtained by annealing VO$_2$ films at
600$^o$C for 30 minutes in a vacuum chamber where the pressure is
maintained in the order of 10$^{-6}$ torr.\cite{yun} The VO$_2$
films are grown by the sol-gel method on (10$\bar{1}$0) oriented
sapphire substrates~\cite{chae} and were used as the starting
material in order to obtain crystalline V$_2$O$_3$ films. The
films were confirmed by X-ray diffraction to be single phase
V$_2$O$_3$. The resistivity of the V$_2$O$_3$ films shows a
decrease of more than four orders of magnitude in the metallic
phase compared to the insulating phase with $T_c$ = 150 K at the
midpoint of the MIT (in the heating cycle) in good agreement with
previous reports.\cite{metcalf} Although the VO$_2$ and V$_2$O$_3$
films are polycrystalline, the MIT characteristics closely
resemble properties of bulk samples.\cite{chae,yun,metcalf}

The optical constants of the VO$_2$ films were obtained from
ellipsometric data (50 meV - 6 eV) and near-normal incidence
reflectance data (5 - 80 meV) in the insulating and metallic
states. The optical constants of V$_2$O$_3$ films were obtained
from a combination of ellipsometric measurements and near-normal
incidence reflectance, with ellipsometric data covering most of
the frequency range. Ellipsometric data was obtained from 0.6 - 6
eV for $T$ = 100 - 400 K, and from 50 meV - 0.7 eV for $T$ = 300 -
400 K. Reflectance data was obtained from 50 meV to 0.7 eV from
$T$ = 100 - 300 K, and from 6 - 80 meV in the temperature range
$T$ = 100 - 400 K. Unlike previous optical studies on single
crystals, the VO$_2$ and V$_2$O$_3$ films were not subject to
surface treatment (polishing etc.) which could affect the spectra
in the visible-ultraviolet spectral range. Therefore the optical
constants are those of the pristine as-grown films.

\begin{figure}[t]
\epsfig{figure=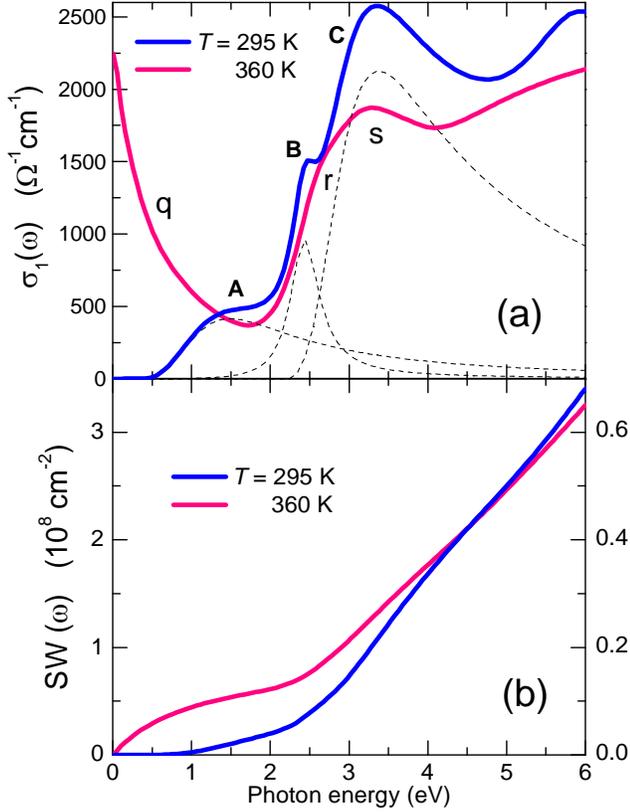,width=90mm,height=110mm}
\caption{(color online):(a) Real part of the conductivity
$\sigma_1$($\omega$) of VO$_2$ plotted as a function of photon
energy for the $M_1$ insulating phase ($T$ = 295 K) and the rutile
metallic phase ($T$ = 360 K). The labels in upper and lower case
letters refer to features in $\sigma_1$($\omega$) in the
insulating phase and metallic phase respectively. The dashed
curves represent the optical transitions that contribute to
features ``A", ``B", and ``C" in $\sigma_1$($\omega$) in the
insulating phase (see text for details). (b) The spectral weight
$SW$($\omega$) is plotted as a function of photon energy for the
insulating and rutile metallic phases of VO$_2$. The spectral
weight is also plotted in units of energy $K$($\omega$) defined in
eq.~\ref{energy}.} \label{sigma1-VO2}
\end{figure}

\begin{figure}[t]
\epsfig{figure=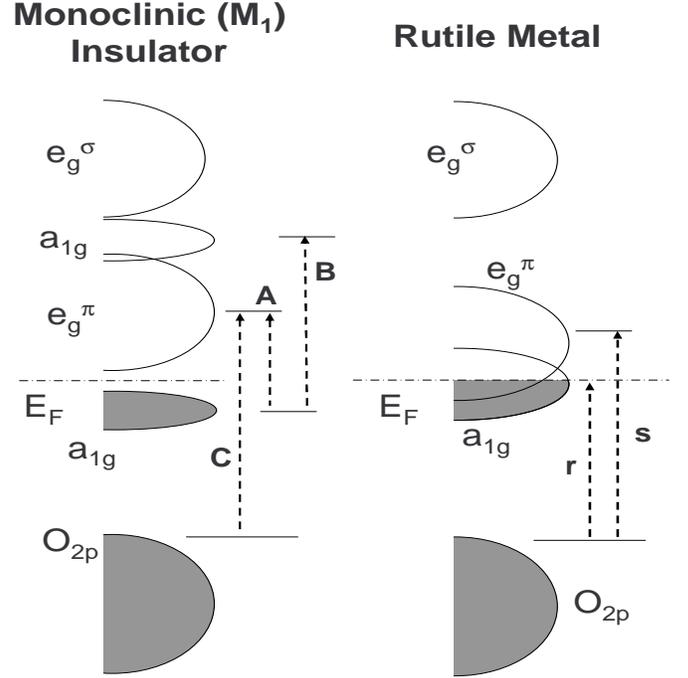,width=90mm,height=95mm}
\caption{Energy level diagrams for the monoclinic insulating
($M_1$) phase and rutile metallic ($R$) phase of VO$_2$ showing
the relevant vanadium and oxygen energy levels and optical
transition energies. $E_F$ denotes the Fermi level. The energy
level diagram in the $M_1$ phase is based on the results of this
work and refs.\onlinecite{xray-photo,eels} and includes the effect
of $U$ on the $t_{2g}$ bands. The energy level diagram of the
rutile metal is based on this work as well as band theory
calculations that give the relative separation between the
$t_{2g}$ ($a_{1g}$, $e_g^{\pi}$), $e_g^{\sigma}$ and O$_{2p}$
bands.~\cite{allenprl,eyert} However, the effect of $U$ on the
$t_{2g}$ bands in the rutile metal is not explicitly included in
band theory calculations and may lead to a Hubbard splitting of
the $a_{1g}$ band.\cite{georges} In this case, the feature ``r" in
Fig.~\ref{sigma1-VO2}a may alternatively arise from an optical
transition between the lower and upper Hubbard bands.}
\label{bands-VO2}
\end{figure}

The real and imaginary parts of the ellipsometric constants of the
VO$_2$ and V$_2$O$_3$ films on sapphire substrates were measured
for two angles of incidence (60$^\circ$ and 75$^\circ$). The
complex optical constants (and subsequently the complex
conductivities) of VO$_2$ and V$_2$O$_3$ were obtained via the
analysis of a two-layer model of a film layer on an infinitely
thick substrate. The latter aspect of the model is valid for a
substrate with rough back-surface which prevents coherent
back-reflections.\cite{tompkins,burch} Details of the
ellipsometric measurements at cryogenic temperatures on V$_2$O$_3$
are given in the Appendix. The electromagnetic response of the
VO$_2$ and V$_2$O$_3$ films is modelled using Drude, Lorentzian
and Tauc-Lorentzian oscillators. The complex optical constants of
the bare sapphire substrate have been obtained via ellipsometric,
near-normal incidence reflectance and normal-incidence
transmission measurements and were incorporated into the
model.\cite{tiwald} The absolute near-normal incidence reflectance
measurements on the VO$_2$ and V$_2$O$_3$ films were performed
using in-situ evaporated gold film as a reference. The reflectance
data was analyzed together with ellipsometric data using the same
two-layer model as described previously. The contributions of
VO$_2$ and V$_2$O$_3$ phonons to the optical constants in the
spectral range $\approx$ 20-100 meV were modelled by Lorentzian
oscillators. Subsequently, these phonon contributions were
subtracted and only the electronic contribution to the optical
constants of VO$_2$ and V$_2$O$_3$ is presented and analyzed in
this work. We note that the spectral weight of the VO$_2$ and
V$_2$O$_3$ phonons is negligibly small compared to the electronic
spectral weight, and therefore we will not be concerned with the
phonon spectral weight in this work.\cite{footnote2}

\section{Results on VO$_2$}

In Fig.~\ref{sigma1-VO2}a, we plot the real part of the optical
conductivity $\sigma_1$($\omega$) of the $M_1$ insulating phase
($T$ = 295 K) and rutile metallic phase ($T$ = 360 K) of a VO$_2$
film. The features in $\sigma_1$($\omega$) of the insulating and
metallic phases will be explained within the model of energy
levels shown in Fig.~\ref{bands-VO2}. According to band theory
calculations and photoemission
data,\cite{allenprl,eyert,shin,okazaki-photo,xray-photo} the
O$_{2p}$ bands lie between 2 eV and 8 eV below the Fermi energy in
both the insulating and metallic phases. Also, within band theory
calculations, the crystal field splits the degenerate $d$-orbitals
into $t_{2g}$ bands and e$_g^\sigma$ bands. The former are lower
in energy and contain the single $d$-electron. The $t_{2g}$ bands
are further split into an $a_{1g}$ band and $e_g^\pi$ bands, with
the latter centered at higher energy compared to the former. We
follow Ref.\onlinecite{georges} in using the terminology $a_{1g}$
and $e_g^\pi$. The $a_{1g}$ band further splits into what we call
the lower $a_{1g}$ band and the upper $a_{1g}$ band in the $M_1$
insulator. In the Mott picture, these can be considered as lower
and upper Hubbard bands while in the Peierls model, these are the
bonding and anti-bonding bands
respectively.\cite{goodenough,mottprb} The energy gap of 0.6 eV in
the monoclinic ($M_1$) insulating phase is between the filled
$a_{1g}$ band and the empty $e_g^\pi$ band. Peierls physics alone
cannot explain this large energy gap as pointed out by
Mott.\cite{mottprb} Various LDA calculations repeatedly fail to
confirm this large energy gap.\cite{allenprl,eyert} The failure of
LDA to account for a gap in insulating VO$_2$ has been attributed
to the shortcomings of LDA in predicting band gaps
correctly.\cite{allenprl} However, an energy gap of correct
magnitude does appear once correlation effects (Hubbard $U$) are
taken into account.\cite{georges,jan} In the rutile metal, the
energy gap collapses and the Fermi level crosses partially filled
$a_{1g}$ and e$_g^\pi$ bands (see Fig.
2).\cite{goodenough,mottprb,allenprl,eyert,georges}

We attempt to explain the various features in the optical spectra
of VO$_2$ (and subsequently of V$_2$O$_3$) by comparing our data
to previous theoretical and experimental reports. Peaks at finite
energies in the optical conductivity occur when there are direct
optical transitions between filled and empty bands in a solid.
More precisely, the interband optical transitions are governed by
the electric dipole transition matrix elements and peaks in the
joint density of (filled and empty) states. The hump feature in
Fig.~\ref{sigma1-VO2}a labelled ``A" at 1.4 eV is due to optical
transitions from the filled lower $a_{1g}$ band to the empty
$e_g^{\pi}$ bands across an optical gap of $\approx$ 0.5 eV. It is
appropriate to attribute peak ``B" at 2.5 eV to transitions from
the narrow lower filled $a_{1g}$ band to the narrow upper empty
$a_{1g}$ band. Peak ``C" at 3.2 eV is due to transitions from the
filled O$_{2p}$ bands and empty $e_g^{\pi}$ bands. We do not
observe any obvious signs of optical transitions from the O$_{2p}$
bands to the empty $a_{1g}$ band which are expected at $\approx$
4.7 eV, and this is likely due to diminished overlap between the
$a_{1g}$ and O$_{2p}$ orbitals in the insulating state of VO$_2$.

A broad Drude-like feature is observed in the optical conductivity
of the rutile metallic phase (labelled ``q"). The spectral weight
in the Drude-like peak is borrowed from the optical transitions at
frequencies higher than 1.3 eV, the energy at which the two solid
curves in Fig.~\ref{sigma1-VO2}a cross. We see a weak shoulder
``r" at 2.6 eV and a peak ``s" at 3.1 eV which we attribute to
optical transitions from the O$_{2p}$ bands to partially filled
$a_{1g}$ and $e_g^{\pi}$ bands respectively. Compared to the
monoclinic insulator, the optical transitions from O$_{2p}$ to the
$a_{1g}$ band become apparent in the rutile metal possibly because
of increased overlap between the O$_{2p}$ and $a_{1g}$ orbitals.
An alternative explanation of the shoulder feature labelled ``r"
is that it arises from optical transitions between remnants of the
lower Hubbard band and upper Hubbard band in the rutile metal. We
note that while the lower Hubbard band is seen by photoemission, a
feature due to the upper Hubbard band is absent in the X-ray
absorption data.\cite{xray-photo}

At this point we remark that compared to the broader peaks ``A"
and ``C" in $\sigma_1$($\omega$) in the insulating phase, peak
``B" is unusually narrow. This is more so in the case of VO$_2$
film grown on (10$\bar{1}$0) sapphire whose optical conductivity
we published in a previous work.\cite{qazilbash} In fact, peak
``B" is described by a nearly Lorentzian lineshape with half-width
of 0.5 eV which is much narrower compared to the corresponding
width parameters of the Tauc-Lorentzian oscillators~\cite{T-L}
used to describe peaks ``A" ($\approx$ 1.6 eV) and ``C" ($\approx$
1.4 eV). The quasi-one-dimensional nature of the V-V chains leads
to narrow upper and lower $a_{1g}$ bands and a sharp optical
transition between these bands. In addition, it is possible that
peak ``B" is excitonic in nature,\cite{verleur} such that the
optical transition between the filled and empty $a_{1g}$ bands is
red-shifted by the electron-hole binding energy due to Coulomb
attraction. Such an electron-hole pair is described as a
charge-transfer exciton~\cite{tsuda} whose spatial extent lies
somewhere between the Mott-Wannier exciton that is delocalized
over several unit cells and the Frenkel exciton that is confined
to an atom or molecule. The charge-transfer exciton is likely
localized on the vanadium dimer whose V-V separation is 2.65 \AA,
and thus is more closely related to the Frenkel exciton. The
charge-transfer exciton is not present in the rutile metallic
phase and this could be due to one or more of the following
reasons: merger of the lower and upper $a_{1g}$ levels, absence of
long-range charge-ordering, and screening due to mobile charge
carriers.

We make use of the model that is applicable to Mott-Wannier
excitons to estimate the binding energy ($E_b$) of the
charge-transfer exciton. It is not obvious if this model is
applicable to the charge-transfer exciton in VO$_2$ because of the
exciton's localized character. However, the model appears to
provide a reasonable description of charge-transfer excitons in
alkali halides and pentacene, for example.\cite{schuster} The
binding energy is given by $E_b=\frac{e^2}{\epsilon{r_{eh}}}$
where $e$ is the electron charge, $\epsilon$ is the static
dielectric constant of VO$_2$ ($\epsilon \approx$ 20) and $r_{eh}$
is the separation between the electron-hole pair in the exciton
and is taken to be 2.65 \AA ~which is the V-V distance in the
dimer. We hence obtain $E_b \approx$ 0.27 eV. Therefore, we
estimate the actual energy separation between the lower and upper
$a_{1g}$ levels to be 2.8 eV. The photoemission and polarized
x-ray absorption data suggest that the separation between the
$a_{1g}$ levels is indeed 2.5 - 2.8 eV.\cite{xray-photo} It
appears that the higher estimate is closer to the picture we
present here. The half-width of the $a_{1g}$ bands in
Ref.\onlinecite{xray-photo} is $\approx$ 1 eV. Interband
transitions between the lower and upper $a_{1g}$ energy levels
would then be expected to be described by a Lorentzian oscillator
with a half-width of at least 1 eV. However, peak ``B" is
described by a Lorentzian oscillator with a half-width of
$\approx$ 0.5 eV which lends support to the hypothesis that it is
excitonic in nature. We note that the assignment of the interband
transitions ``A", ``C", ``r" and ``s" are also consistent with the
photoemission and x-ray absorption data.\cite{xray-photo}

Calculations for the separation of $a_{1g}$ levels in the V-V
dimer start with considering a ``hydrogen molecule" model of the
dimer.~\cite{jan} With the intra-atomic Coulomb repulsion $U$ set
to 4.0 eV and the LDA hopping ($t$) to 0.7 eV, the
$a_{1g}$-$a_{1g}$ separation is given by $\Delta_{a_{1g}} = -2t +
4t\sqrt{1 + (U/4t)^2} = $ 3.48 eV. This value of $U$ is required
to open an energy gap of the correct magnitude in insulating
VO$_2$. The calculated value for $\Delta_{a_{1g}}$ is expected to
be reduced to 3 eV due to effects not accounted for by the above
simple model and discussed in Ref.\onlinecite{jan}. The actual
position of peak ``B" at 2.5 eV is less than the theory value of
the $a_{1g}$-$a_{1g}$ energy separation. This further supports the
notion of its excitonic origin due to final state effects of
electron-hole Coulomb attraction that are not accounted for in
Ref.\onlinecite{jan}.

We now discuss the spectral weight changes across the MIT in
VO$_2$. The oscillator strength sum-rule (or $f$-sum rule) is a
fundamental statement on the conservation of charge in a material.
The most generic form employs the integral of the real part of the
optical conductivity $\sigma_1$($\omega$) over all frequencies:

\begin{equation}
\int_{0}^{\infty}{\sigma_1(\omega)}{d\omega}=\frac{{\pi}ne^2}{2m_e}
\label{sumrule}
\end{equation}

Here, $n$ is the density of electrons and $m_e$ is the free
electron mass. In reality, however, we study spectral weight
changes up to a finite frequency, which in our experiments is 6
eV, and we can comment upon whether or not the $f$-sum rule is
satisfied in this frequency window. We plot the spectral weight
$SW(\omega)=\int_{0}^{\omega}{\sigma_1(\omega')}{d\omega'}$ in
Fig.~\ref{sigma1-VO2}b for VO$_2$. The right-hand vertical axis in
Fig. 1b shows the spectral weight as an energy $K$($\omega$) in
units of eV which can readily be compared to band theory values.
Following Millis \emph{et al.}, the energy $K$($\omega$) is given
by:\cite{millisndoped}

\begin{equation}
K(\omega)=\frac{{\hbar}a}{e^2}\int_{0}^{\omega}\frac{2\hbar}{\pi}{\sigma_1(\omega'}){d\omega'}
\label{energy}
\end{equation}

Here $a$ is the lattice constant which is taken as $\approx$ 3
\AA, the average V-V distance in both VO$_2$ and V$_2$O$_3$. Also
note that $\hbar/e^2 =$ 4.1 k$\Omega$.

Fig.~\ref{sigma1-VO2}a,b show that the spectral weight in the
broad Drude-like part of the conductivity in the rutile metal is
obtained from the loss of spectral weight of the peaks ``A", ``B"
and ``C" in the insulator. In the rutile metal, the $e_g^{\pi}$
bands and the $a_{1g}$ band are both partially occupied (see
Fig.~\ref{bands-VO2}). Therefore, the optical transition (peak
``s") from the O$_{2p}$ to the $e_g^{\pi}$ band has reduced
spectral weight in the rutile metal compared peak ``C" in the
insulator. The changes in spectral weight across the MIT extend up
to and beyond 6 eV. We note that interband transitions between the
filled O$_{2p}$ and empty $e_g^{\sigma}$ levels may occur above 4
eV but are not expected to change across the MIT. Interband
transitions between filled $t_{2g}$ and empty $e_g^{\sigma}$ are
expected at $\approx$ 6 eV ($U$ $\approx$ 4 eV in addition to 2 eV
crystal field splitting)~\cite{lee} and are likely to be modified
by the MIT because of the rearrangement of $t_{2g}$ bands. Thus
the changes in optical conductivity and spectral weight that occur
above 4 eV across the MIT can be attributed to changes in optical
transition probabilities from filled $t_{2g}$ states to empty
$e_g^{\sigma}$ states. The rather large energy scale over which
changes in $SW$($\omega$) are seen across the metal-insulator
transition ought to be taken as direct evidence for the
predominance of correlation effects, and hence an indication of a
Mott transition. Similar large energy scales are involved in
spectral weight changes in doped Mott insulators, for example, in
the cuprates.~\cite{imada-rmp,basovreview} We find that the
Hubbard parameter $U$ determines, to a large extent, the energy
scale over which $SW$($\omega$) changes take place across the MIT
in VO$_2$. The Peierls instability within the single-particle
scenario alone would result in rearrangement of the $t_{2g}$ bands
within $\approx$ 1 eV of the Fermi energy. Then the scale over
which changes in spectral weight occur would be $\approx$ 3 eV and
determined by the interband transitions between O$_{2p}$ and
$t_{2g}$ bands and between $t_{2g}$ and $e_g^{\sigma}$ bands
(without Hubbard $U$). Thus changes in $\sigma_1$($\omega$) and
$SW$($\omega$) that occur well beyond 3 eV provide direct evidence
of the importance of correlation effects (Hubbard $U$) to the MIT
in VO$_2$.

The energy $K$($\omega$) associated with the intraband
conductivity (the broad Drude-like feature) in VO$_2$ is $\approx$
0.12 eV and can be considered as the kinetic energy of delocalized
charge carriers.\cite{millisreview} The kinetic energy is
determined by setting the intraband cutoff energy at 1.7 eV where
the minimum in $\sigma_1$ ($\omega$) occurs. The kinetic energy of
the delocalized carriers is nearly fifty percent of the band
theory value.\cite{allenprl,kotliar,qazilbash} Correlations
amongst electrons are expected to reduce their kinetic energy
compared to that predicted by band
theory.\cite{millisndoped,millisreview} On the other hand, in the
presence of electron-phonon interactions, the ground state
wavefunction and hence the kinetic energy of the electrons is
essentially that given by band theory.\cite{millisreview} An
exception to this occurs in the presence of very strong
electron-phonon coupling which can cause a significant reduction
in the electronic kinetic energy and likely leads to a polaronic
insulating state. However, this scenario is not realized in the
rutile phase of VO$_2$ which is clearly metallic. Therefore, the
reduction in kinetic energy of the electrons clearly points to the
dominance of correlation effects to the charge dynamics in rutile
metallic VO$_2$. In addition, it was shown in Ref.
\onlinecite{qazilbash} that the form of the frequency-dependent
scattering rate cannot be explained by electron-phonon scattering
alone, and that electronic correlations must also influence the
charge dynamics. More recently, divergent optical mass was
inferred in the metallic nano-scale islands that form in the
insulating host at the onset of the MIT in
VO$_2$.\cite{massdivergence} This observation is strong evidence
for the importance of Mott physics to the insulating and metallic
phases in the vicinity of the MIT in VO$_2$.

\begin{figure}[t]
\epsfig{figure=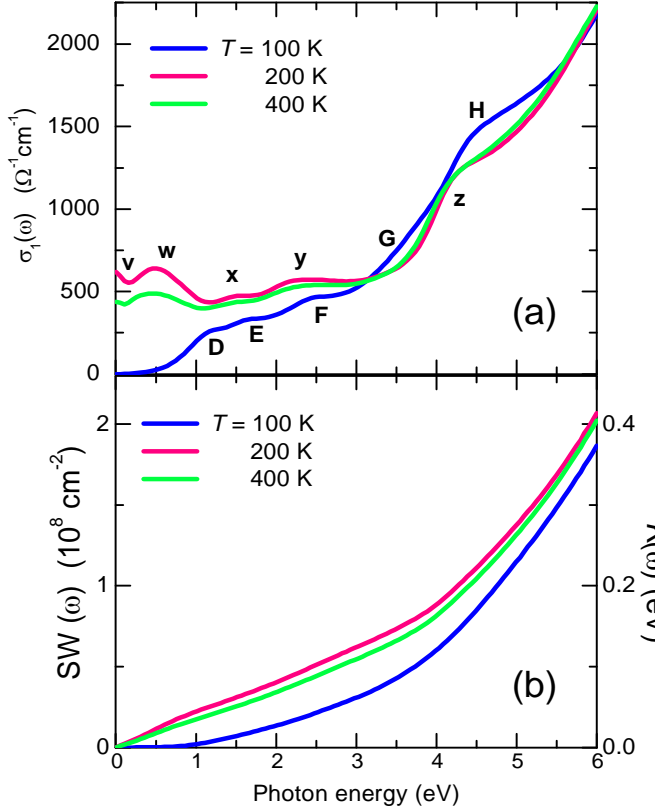,width=90mm,height=110mm}
\caption{(color online): Real part of the conductivity
$\sigma_1$($\omega$) of V$_2$O$_3$ plotted as function of photon
energy for the insulating phase ($T$ = 100 K) and for two
temperatures in the metallic phase ($T$ = 200 K and 400 K). The
labels in upper and lower case letters refer to features in
$\sigma_1$($\omega$) in the insulating phase and metallic phase
respectively. (b) The spectral weight $SW$($\omega$) is plotted as
a function of photon energy for selected temperatures. The
spectral weight is also plotted in units of energy $K$($\omega$)
defined in eq.~\ref{energy}.} \label{sigma1-V2O3}
\end{figure}

\begin{figure}[t]
\epsfig{figure=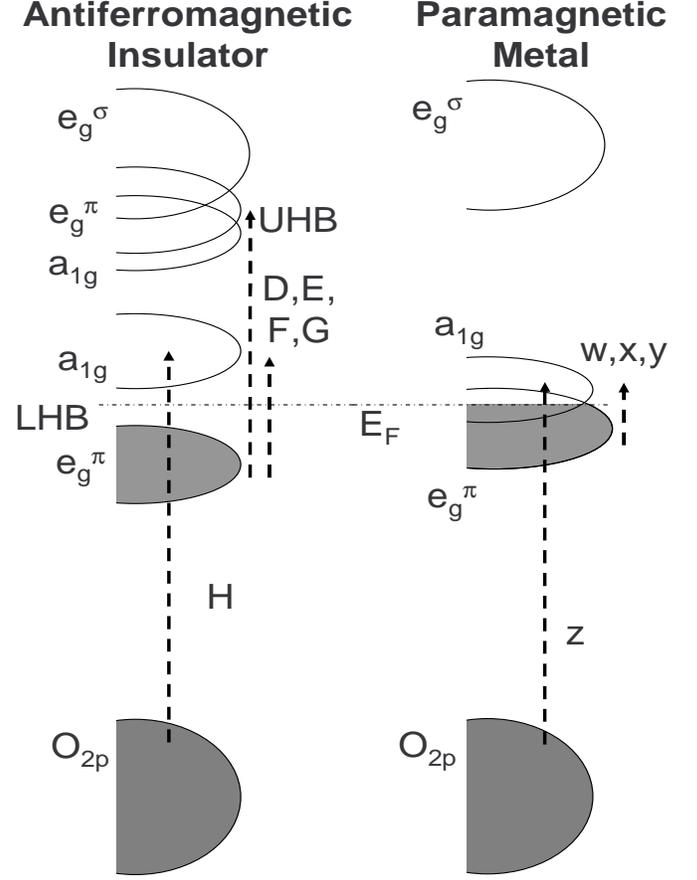,width=90mm,height=120mm}
\caption{Energy level diagrams for the antiferromagnetic
insulating (AFI) phase and paramagnetic metallic (PMM) phase of
V$_2$O$_3$ showing the relevant vanadium and oxygen energy levels
and optical transitions. LHB and UHB refer to the lower Hubbard
bands and upper Hubbard bands respectively. $E_F$ denotes the
Fermi level. The energy level diagram for the AFI is based on the
LDA+U calculations\cite{ezhov}. For the PMM energy levels we rely
on LDA calculations,\cite{mattheiss} but we note that the
persistence of interband transitions between the vanadium $t_{2g}$
orbitals in $\sigma_1$($\omega$) in the PMM suggests the
importance of the Hubbard $U$ which is not accounted for within
LDA. Our work and ref.~\onlinecite{shin} confirm the relative
separation of the O$_{2p}$ and vanadium $t_{2g}$ ($a_{1g}$ and
$e_g^{\pi}$) bands.}\label{bands-V2O3}
\end{figure}

\section{Results on V$_2$O$_3$}

We now turn to the real part of optical conductivity
$\sigma_1$($\omega$) of V$_2$O$_3$ plotted in
Fig.~\ref{sigma1-V2O3}a. As in VO$_2$, there is rearrangement of
$\sigma_1$($\omega$) and $SW$($\omega$) across the MIT. In the
insulating phase of V$_2$O$_3$ at $T$ = 100 K, there is an optical
gap of $\approx$ 0.5 eV followed by interband transitions labelled
D to H. Because there are two valence $d$-electrons per vanadium
ion in V$_2$O$_3$, V-V intersite transitions are expected at
multiple energies and hence multiple peaks occur in the
conductivity data up to 3.5 eV.~\cite{lee} The energy band diagram
for V$_2$O$_3$ in Fig.~\ref{bands-V2O3} gives the relative energy
separation between the O$_{2p}$ and vanadium $t_{2g}$ bands and is
a rough guide for the interband transitions and the changes that
take place across the MIT. The photoemission data and band
structure calculations show that the O$_{2p}$ bands lie between
3.5 eV and 8.5 eV below the Fermi energy.\cite{shin,mattheiss}
Then the rise in optical conductivity starting from $\approx$ 4 eV
is due to interband transitions from the O$_{2p}$ to the empty
$t_{2g}$ orbitals on the vanadium sites. Specifically, the
distinct shoulder-like structure ``H" in the AFI spectrum can be
attributed to such interband transitions. The crystal field splits
the vanadium $d$-bands into $e_g^{\sigma}$ bands that are centered
$\approx$ 3 eV above the $t_{2g}$ bands.\cite{ezhov,mattheiss} The
lower symmetry of the rhombohedral lattice lifts the degeneracy of
the $t_{2g}$ bands, which evolve into non-degenerate $a_{1g}$ and
doubly degenerate $e_g^{\pi}$ bands.\cite{ezhov}

We focus our attention on the optical transitions within the
vanadium $t_{2g}$ manifold which are better described by
considering intersite transitions in real space rather than by
appealing to energy bands. This is because in V$_2$O$_3$, two
$d$-electrons are believed to be distributed in the $e_g^{\pi}$
and $a_{1g}$ bands in a non-trivial
manner.\cite{castellani,ezhov,mila,joshi,matteo,tanaka,bao,paolasini,
park,mo} Lee $\emph{et al.}$ have discussed, on general physical
grounds, the optical excitations in transition metal oxides with
multi-orbital character.\cite{lee} However, their classification
scheme does not take into account explicitly the lifting of
degeneracy of the $t_{2g}$ orbitals due to crystal field splitting
which may be relevant for identification of the inter-site
transitions in V$_2$O$_3$. Therefore, assignment of the various
interband transitions in both the AFI and PMM phases is difficult
without a quantitative model specific to V$_2$O$_3$. Nevertheless,
relevant physical parameters for V$_2$O$_3$ can be estimated from
our data based on the classification scheme of Lee $\emph{et al}$.
We note that for two valence electrons in the $t_{2g}$ manifold,
the minimum and maximum allowed optical transition energies are
given by $U - 3J_H$ and $U + 2J_H$ respectively while assuming the
crystal field splitting is less than the other relevant parameters
in the system ($U$, $J_H$ and bandwidth). Here, $U$ is the
intra-atomic Coulomb repulsion and $J_H$ is the Hund's rule
exchange energy. Then the lowest energy peak ``D" at 1.2 eV in the
AFI spectrum is the optical transition given by $U - 3J_H$ and the
energy peak ``G" at 3.6 eV is given by $U + 2J_H$. This gives $U$
$\approx$ 2.6 eV and $J_H$ $\approx$ 0.5 eV. These numbers are
towards the lower end of the spectrum of values that appear in the
extensive literature on V$_2$O$_3$ (for example, see
Ref.~\onlinecite{matteo}). It is likely that we have estimated the
screened value of $U$ while the unscreened value could be 4-5
eV.~\cite{ezhov} We note that in the antiferromagnetic insulator,
each vanadium ion has one nearest neighbor with spin aligned in
the same direction and two nearest neighbors with spins
anti-aligned. Also, because there are no definitive reports on
orbital ordering in AFI, all allowed optical transitions should
appear in the spectra of the AFI.\cite{lee} We also note here that
for reasons given earlier for VO$_2$, the transitions from
O$_{2p}$ and $t_{2g}$ bands to the $e_g^{\sigma}$ bands lie close
to and beyond the energy cutoff of 6 eV in our
experiment.\cite{lee} The spectral features due to inter-site
transitions between the $t_{2g}$ orbitals of the vanadium ions in
V$_2$O$_3$ closely resemble those seen in the Mott insulator
YVO$_3$ in which the vanadium ions also have two $d$-electrons in
the $t_{2g}$ orbitals.\cite{vdmarel} In YVO$_3$, the spectral
features between 1 eV and 4 eV were also assigned to inter-site
transitions between the $t_{2g}$ orbitals.

From Fig.~\ref{sigma1-V2O3}a,b one can see that there is
significant rearrangement of the conductivity spectra across the
MIT. A weak Drude peak labeled ``v" and a finite energy mode at
0.5 eV labeled ``w" emerge in the metallic phase together with
spectral weight enhancement of other interband transitions below 3
eV. In the PMM state, the spectral weight due to the shoulder-like
features ``G" and ``H" decreases. The peaks ``E", ``F" and ``H" in
the AFI spectrum shift downwards by ~ 0.2-0.3 eV and are labelled
as ``x", ``y" and ``z" respectively in the spectra for PMM. This
downward shift could be due to the reduction of crystal field
splitting between the $a_{1g}$ and $e_g^{\pi}$ levels. The reduced
spectral weight of the ``z" feature in PMM compared to ``H"
feature in AFI is due to the partial occupation of the $a_{1g}$
band in the PMM phase. The decrease in the spectral weight of the
$U + 2J_H$ transition (feature ``G") is due to the absence of
antiferromagnetic ordering in PMM. Moreover, it appears that
ferromagnetic correlations (not ordering) persist in the PMM
because of the increase in the optical conductivity (and spectral
weight) between 1 and 3 eV compared to the AFI. This is because
for ferromagnetic correlations, the optical transitions between
$U$ and $U - 3J_H$ will be favored compared to those between $U$
and $U + 2J_H$.\cite{bao,taylor,khomskii}. Moreover, orbital
switching effects could also contribute to the increased optical
conductivity between 1 and 3 eV in the PMM.\cite{khomskii}

The Drude peak in the PMM arises from intraband dynamics of
coherent quasiparticles of $a_{1g}$ character. This is firmly
supported by DMFT calculations in the PMM state.\cite{biermann}
Since the spectral weight of the Drude peak is low, we surmise
that the density of coherent quasiparticles is rather low and/or
their mass is quite high. Nevertheless the presence of the Drude
peak in metallic V$_2$O$_3$ with $dc$ conductivity below that of
the apparent Ioffe-Regel-Mott limit (4000 $\Omega^{-1}$cm$^{-1}$)
of metallic transport is remarkable.\cite{qazilbash} Next, it
appears that the majority of the charge carriers are completely
incoherent and reside in the $e_g^{\pi}$ bands.\cite{biermann}
These likely contribute a broad background to the conductivity and
do not give rise to any particular feature in the PMM optical
conductivity. The charge dynamics in the $dc$ limit is then
dominated by the coherent $a_{1g}$ quasiparticles. The peak ``w"
can be assigned to intersite transitions between incoherent
$e_g^{\pi}$ carriers and partially occupied coherent $a_{1g}$
orbitals.

In Fig.~\ref{sigma1-V2O3}b, we see that there is more spectral
weight in the metallic phase ($T$ = 200K) at low energies compared
to the insulating phase ($T$ = 100K) and that this extra spectral
weight is never fully recovered in the insulating phase up to 6
eV. The additional spectral weight in PMM up to 3 eV is partly
borrowed from optical transitions ``G" and ``H" in AFI. The rest
is probably borrowed from higher-lying optical transitions ($>$ 6
eV) in the AFI between the O$_{2p}$ bands and the upper Hubbard
bands as well as between $e_g^{\pi}$ and $e_g^{\sigma}$ bands.
More interesting is the increase in spectral weight at low
energies (up to 3 eV) upon cooling metallic V$_2$O$_3$ from $T$ =
400K to $T$ = 200 K, in accord with previous
reports.\cite{rozenbergprl,rozenbergprb} However, in contrast to
previous reports, our data extends up to much higher energies (6
eV) and shows that this increase in spectral weight is only
partially accounted for by a corresponding decrease between 3 eV
and 6 eV. This is very different from what is expected in cooling
a conventional metal that is well-described by band theory. In a
conventional metal, the Drude peak narrows as the scattering rate
decreases with temperature but the spectral weight is conserved
within the bandwidth. In V$_2$O$_3$, the total $t_{2g}$ bandwidth
according to band theory is $\approx$ 2.5 eV.\cite{mattheiss}
Therefore, the temperature-driven spectral weight changes in
metallic V$_2$O$_3$ that span energies at least up to 6 eV cannot
be simply explained within a single-particle picture.

\begin{figure}[t]
\epsfig{figure=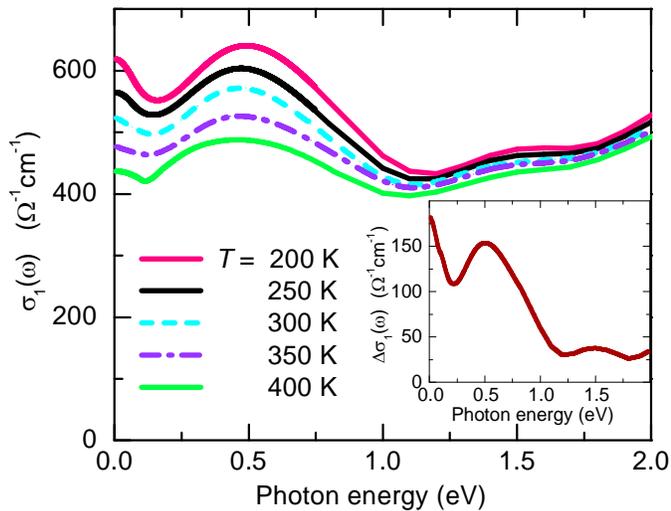,width=90mm,height=70mm}
\caption{(color online): Temperature dependence of the real part
of the optical conductivity $\sigma_1$($\omega$) of the metallic
phase (PMM) of V$_2$O$_3$ plotted as a function of the incident
photon energy up to 2 eV. The inset shows the difference in the
optical conductivities of PMM between the lowest and highest
temperatures $\Delta\sigma_1$($\omega$)= $\sigma_1$($\omega$, $T$
= 200 K) $-$ $\sigma_1$($\omega$, $T$ = 400 K).}
\label{sigma1-T-V2O3}
\end{figure}

In the PMM phase of V$_2$O$_3$, the energy $K$($\omega$) does not
reach the intra-band value of VO$_2$ (0.12 eV) until about 3 eV of
the incident photon energy, well into the region of V-V inter-band
transitions. If we assume that only the narrow Drude peak in PMM
V$_2$O$_3$ represents the intra-band part of the energy and
spectral weight, then we obtain $K$($\omega$) $\approx$ 0.012 eV,
the kinetic energy of delocalized carriers, which is nearly six
percent of the band value in V$_2$O$_3$~\cite{rozenbergprb} and
significantly less than in VO$_2$. This will place PMM V$_2$O$_3$
in the strongly correlated limit, with a higher degree of
correlations compared to rutile VO$_2$. Therefore, correlation
effects are more pronounced in the metallic phase (PMM) near a
magnetically-ordered Mott insulator (AFI V$_2$O$_3$) compared to
the rutile metallic phase ($R$) near the Mott insulator ($M_1$
VO$_2$) that is charge-ordered with quasi-one-dimensional chains
of vanadium pairs. Nevertheless, we emphasize that charge dynamics
of rutile metallic VO$_2$ cannot be described without taking into
account correlation effects.\cite{qazilbash}

In Fig.~\ref{sigma1-T-V2O3} we plot the temperature dependence of
the low energy optical conductivity of the PMM. The difference
between the $T$ = 200K and $T$ = 400 K optical conductivity is
plotted in the inset of Fig.~\ref{sigma1-T-V2O3}. The
temperature-induced change in the conductivity for the peak at 0.5
eV in the PMM is very similar to that of the Drude peak. This
supports our assignments of the Drude peak to the coherent
quasiparticles in $a_{1g}$ and the peak at 0.5 eV to inter-site
transitions between the incoherent carriers in $e_g^{\pi}$ and
coherent empty states of $a_{1g}$ character. This is because the
coherent quasiparticles are expected to show a strong temperature
dependence compared to the charge carriers in $e_g^{\pi}$ that are
well above their coherence temperature.\cite{biermann}

\section{Summary and outlook}

We have investigated the optical constants of VO$_2$ and
V$_2$O$_3$ films over a wide frequency and temperature range. We
tracked changes of the optical conductivity $\sigma_1$($\omega$)
and spectral weight $SW$($\omega$) across the metal-insulator
transitions in both materials. In both VO$_2$ and V$_2$O$_3$, the
changes in $\sigma_1$($\omega$) and $SW$($\omega$) extend up to
and beyond 6 eV indicating the importance of electronic
correlations to the metal-insulator transitions. From the kinetic
energy of charge carriers, we deduce that rutile metallic VO$_2$
and paramagnetic metallic V$_2$O$_3$ are correlated metals, with
metallic V$_2$O$_3$ showing a higher degree of correlations. Our
optics data shows that the energy gaps are nearly the same
magnitude in $M_1$ insulating VO$_2$ and anti-ferromagnetic
insulating V$_2$O$_3$ which is interesting given the different
lattice structures, magnetic properties and number of
$d$-electrons in the two oxides. While it may be a coincidence
that the energy gaps have nearly the same magnitude, it is more
likely that the Hubbard $U$ is the common primary factor in
determining the energy gap magnitude in both these oxides. Charge
ordering of vanadium ions imparts a quasi-one-dimensional
character to insulating VO$_2$ resulting in narrow $a_{1g}$ bands
and sharp optical transitions between these bands with possible
excitonic effects.

\begin{figure}[t]
\epsfig{figure=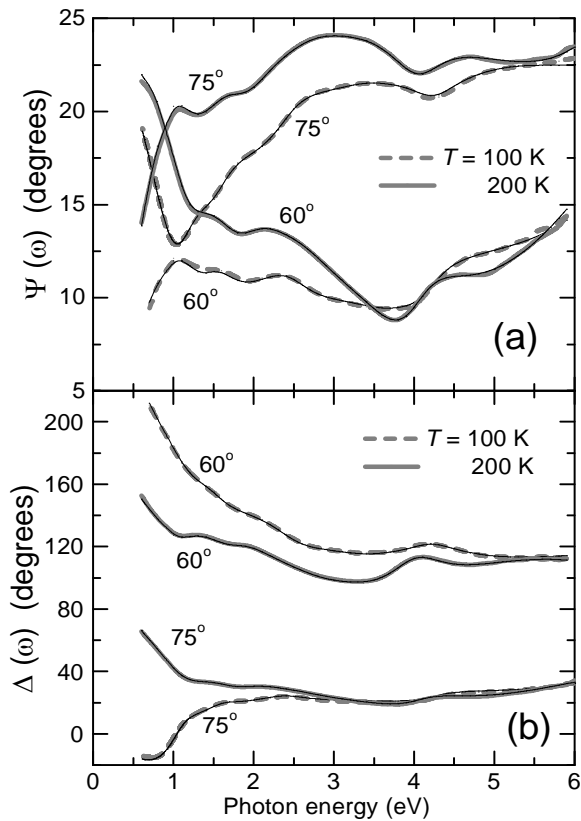,width=80mm,height=110mm}
\caption{The ellipsometric constants $\Psi$($\omega$) and
$\Delta$($\omega$) for the V$_2$O$_3$ film/sapphire substrate
system are plotted as a function of photon energy in panels (a)
and (b) respectively. The thick dashed gray curves are the data
for the AFI phase of V$_2$O$_3$ at $T$ = 100 K for two angles of
incidence (60$^o$ and 75$^o$). The thick solid gray curves are the
corresponding data for the PMM phase of V$_2$O$_3$ at $T$ = 200 K.
The thin solid black curves are fits to the data based on the
model described in the section on experimental methods.}
\label{psidelta}
\end{figure}

\begin{figure}[t]
\epsfig{figure=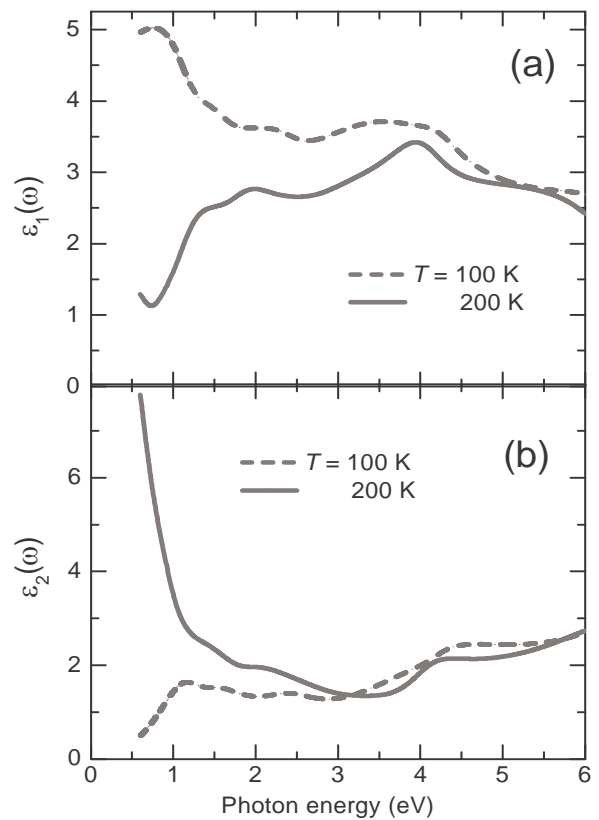,width=80mm,height=110mm}
\caption{The real part $\varepsilon_1$($\omega$) and imaginary
part $\varepsilon_2$($\omega$) of the optical constants for the
V$_2$O$_3$ film are plotted as a function of photon energy in
panels (a) and (b) respectively. The optical constants of
V$_2$O$_3$ film plotted in this figure are obtained from modelling
of the ellipsometric constants of the film-substrate system
displayed in Fig~\ref{psidelta}a,b. The dashed curves are optical
constants of the AFI phase of V$_2$O$_3$ at $T$ = 100 K. The solid
curves are the optical constants for the PMM phase of V$_2$O$_3$
at $T$ = 200 K.} \label{e1e2}
\end{figure}

When electron-phonon interactions cause metal-insulator
transitions to occur, as for example in V$_3$O$_5$ and
Fe$_3$O$_4$, the spectral weight across the phase transitions is
conserved on the energy scale of 1 eV.\cite{lupi2,gasparov} For
the MIT in VO$_2$ and V$_2$O$_3$, the sum rule is not exhausted up
to a much higher energy scale of 6 eV, which clearly highlights
the important role of electron-electron correlations in the MIT in
both these vanadium oxides. Temperature induced changes in optical
conductivity and spectral weight are seen over similar large
energy scales in other transition metal oxides that are Mott
insulators, for example, YVO$_3$, LaVO$_3$ and
LaMnO$_3$.\cite{vdmarel,tokura,keimer,dordevic} Doped Mott
insulators like the cuprates also exhibit shift of spectral weight
from similar high energies to lower energies upon chemical
doping.\cite{imada-rmp,basovreview,dordevic} In this context, our
work indicates that Mott physics plays the dominant role in the
MIT in VO$_2$ and V$_2$O$_3$ whereas electron-phonon interactions
and the lattice rearrangement play a competing role. Nevertheless,
a complete picture of the MIT should take into account the
electron-lattice interactions in addition to electron-electron
interactions.\cite{pfalzer}

Recent work on VO$_2$ provides evidence that the MIT is a Mott
transition driven by Brinkman-Rice mass divergence of the
quasiparticles.\cite{massdivergence,brinkman} The divergent
effective mass occurs in metallic nano-scale puddles that first
nucleate in the insulating host in the vicinity of the
insulator-to-metal transition. The metallic nano-puddles were
observed directly by scanning near-field infrared miscroscopy.
This new data in combination with far-field infrared spectroscopy
measurements demonstrated that the effective mass in the metallic
nano-puddles of VO$_2$ is significantly enhanced compared to the
macroscopic rutile metal at higher temperatures. There is evidence
of heavy quasiparticles in metallic V$_2$O$_3$ in the literature
although mass divergence in V$_2$O$_3$ has not been explicitly
demonstrated.\cite{thomasjltp,mcwhan} Therefore, V$_2$O$_3$ needs
to be studied using the experimental techniques and analytical
procedures employed on VO$_2$ in Ref.\onlinecite{massdivergence}
because it is also possible that in V$_2$O$_3$ mass divergence
occurs in the vicinity of the MIT where there is a tendency
towards phase separation.

In this work, we have displayed evidence of the importance of
correlation effects to the metal-insulator transitions in VO$_2$
and V$_2$O$_3$. A corollary is that structural changes in both
materials cannot explain the large energy gaps in the insulating
phases and the large energy scales that are associated with the
metal-insulator transitions. Nevertheless, structural changes
cannot be easily explained within Mott physics and likely involve
electron-phonon coupling which should be taken into account in any
realistic description of the metal-insulator transitions.
Moreover, the metal-insulator transition temperatures in both the
vanadium oxides we have investigated are an order of magnitude
less than the energy gaps in the insulating phases, and there is
hardly any satisfactory quantitative explanation for this.
However, there are qualitative suggestions emphasizing the
importance of the multi-orbital character of these vanadium
oxides.\cite{khomskii} Future optical/infrared work on single
crystals and/or highly-oriented, epitaxial films of VO$_2$ and
V$_2$O$_3$ is required to determine changes in orbital occupation
that accompany the metal-insulator transition. Nevertheless, the
comprehensive nature of our experiments reported here sets the
stage for more in-depth theoretical analysis and specifically
computational analysis of the metal-insulator transitions since
computational (DMFT and GW) results can be directly compared to
optics data.

\textbf{Note added in proof:} A manuscript by L. Baldassarre
\emph{et al.} on infrared measurements on V$_2$O$_3$ crystals has
been posted recently on the arXiv (arXiv:0710.1247). In this
manuscript, the Drude part of the conductivity of paramagnetic
metallic V$_2$O$_3$ crystals is higher than that of the V$_2$O$_3$
films studied in our work.

\section*{Acknowledgements}

We gratefully acknowledge discussions with S. Biermann, J.
Tomczak, A. J. Millis, G. Kotliar, M. Fogler, J. E. Hirsch, T.
Tiwald and D. van der Marel. This work was supported by Department
of Energy Grant No.DE-FG03-00ER45799 and by ETRI.

\section*{APPENDIX}

For ellipsometry at cryogenic temperatures, the sample has to be
kept in ultra-high vacuum (UHV) to prevent formation of a layer of
ice. The ellipsometric measurements (incident photon energies 0.6
- 6 eV) on the V$_2$O$_3$ film at low temperatures (100 K $\leq T
\leq$ 300 K) were performed in vacuum of $\approx$ 10$^{-9}$ torr
in a custom-built UHV chamber. The incident and reflected light
pass through UHV compatible optical windows. The change in the
polarization state of light due to the windows was accounted for
via calibration in vacuum using a standard silicon wafer with a
200 \AA~thick layer of SiO$_2$.

The ellipsometric constants $\Psi$($\omega$) and
$\Delta$($\omega$) of the V$_2$O$_3$ film on a sapphire substrate
are plotted in Fig.~\ref{psidelta}a,b for the V$_2$O$_3$ film in
the AFI phase ($T$ = 100 K) and in the PMM phase ($T$ = 200 K).
The data presented were obtained for two angles of incidence
(60$^\circ$ and 75$^\circ$). Also plotted are fits to the data
based on the model described in the section on experimental
methods. The optical constants of the V$_2$O$_3$ film for the AFI
and PMM phases are plotted in Fig.~\ref{e1e2}a,b and are obtained
from modelling of the ellipsometric constants of the
film-substrate system.


\begin{references}
\bibitem{footnote1} In contrast to VO$_2$ films grown on SiO$_2$/Si substrates and studied by
X-ray absorption spectroscopy in ref.\onlinecite{xrayfilm}, our
VO$_2$ and V$_2$O$_3$ films grown on lattice matched sapphire
substrates do not show signs of deterioration after cycling
several times through the MIT. This was verified by ellipsometric
and reflectance measurements that give reproducible data in the
insulating and metallic phases before and after cycling through
the MIT.
\bibitem{footnote2} We have studied thin films of VO$_2$ and V$_2$O$_3$
in which the phonon features are fairly weak and are partially
obscured by much stronger phonon features from the sapphire
substrate. Infrared studies of single crystals of VO$_2$ and
V$_2$O$_3$ are required to precisely quantify the phonon
contributions to the optical conductivity and the changes that
occur in the phonons across the MIT.
\end{references}
\end{document}